\documentclass[prb,twocolumn,showpacs,preprintnumbers,amsmath,amssymb]{revtex4} 
 
\usepackage{graphicx}
\usepackage{dcolumn}
 
\newcommand{\beq}{\begin{equation}} 
\newcommand{\eeq}{\end{equation}} 
\newcommand{\beqa}{\begin{eqnarray}} 
\newcommand{\eeqa}{\end{eqnarray}}

\def\tJ{$t$--$J$\ }
\def\ra{\rightarrow}

\def\ts{\textstyle}
\def\nn{\nonumber}
\def\dw{$d$--wave\ }

\def\hc{\mathrm{h.c.}}
\def\kk{\mathbf{k}}
\def\KK{\mathbf{K}}

\def\ee{\mathbf{e}}

\def\iom{{i\omega_n}}
\def\cl{\mathcal{C}}
\def\ord{\mathcal{O}}

\def\G{\mathcal{G}}

\def\teff{t_{\mathrm{eff}}}
\def\flat{\mathrm{flat}}
\def\DCA{\mathrm{DCA}}

\def\bulk{\mathrm{bulk}}

\def\ph#1{\phantom{#1}}
\def\eq#1{(\ref{#1})}
\def\ceq#1{Eq.~(\ref{#1})}
\def\eqs#1#2{(\ref{#1}--\ref{#2})}
\def\ceqs#1#2{Eqs.~(\ref{#1}--\ref{#2})}

\begin{document} 
\title{Cluster Dynamical Mean-Field Methods for d-wave Superconductors: the Role of Geometry} 
 
\author{A. Isidori$^{1}$ and M. Capone$^{2,1}$}

\affiliation{$^{1}$ Dipartimento di Fisica, Universit\`a di Roma ``La Sapienza'', Piazzale A. Moro 2, 00185, Rome, Italy} 
\affiliation{$^{2}$ SMC, CNR-INFM and ISC-CNR, Piazzale A. Moro 2, 00185, Rome, Italy} 

\begin{abstract} 
We compare the accuracy of two cluster extensions of Dynamical Mean-Field Theory in describing $d$-wave superconductors, using as a reference model a saddle-point t-J model which  can be solved exactly in the thermodynamic limit and at the same time reasonably describes the properties of high-temperature superconductors. 
The two methods are Cellular Dynamical Mean-Field Theory, which is based on a real-space perspective, and Dynamical Cluster Approximation, which enforces a momentum-space picture by imposing periodic boundary conditions on the cluster, as opposed to the open boundary conditions of the first method.
We consider the scaling of the methods for large cluster size, but we also focus on the behavior for small clusters, such as those accessible by means of present techniques, with particular emphasis on the geometrical structure, which is definitely a relevant issue in small clusters.
\end{abstract}
\pacs{71.10.Fd, 71.27.+a, 75.20.Hr, 75.10.Lp} 
 
\date{\today} 
\maketitle 

Strongly Correlated electronic materials and the solution of the models introduced to understand their behavior are one of the main challenges of modern solid state physics. Despite the intensive activity triggered by the role of correlations in high-temperature superconductors, many questions remain unanswered.
The vitality of the field is testified by the development of theoretical approaches designed precisely for these systems. Among these method, a central role is played by Dynamical Mean-Field Theory (DMFT)\cite{revdmft}, a non perturbative approach which generalizes the classical mean-field theory to the quantum dynamical world.
The development of DMFT has allowed for a number of successes, starting from  the first unified scenario for the longstanding problem of the Mott transition, and including reliable description of the electronic properties of many correlated systems\cite{revdmft_realistic}.

The idea behind DMFT is analogous to classical mean-field theory, namely
the assumption that each lattice site is equivalent to any other. The
difference with the static case is that, within DMFT, each lattice site
has a completely non-trivial dynamics. Following the above strategy, one
given site can be taken as representative of the whole system: the
lattice problem is therefore mapped onto a dynamical local problem, and
consequently onto a single-impurity model. 
The effect of all the remaining lattice sites will be described by a
bath, whose frequency dependence will be determined through a self-consistency
condition 
that enforces the equivalence between the lattice and the local problems\cite{revdmft,georgeskotliar}.
The main limitation of the standard (single-site)
DMFT, as we have described it so far, is the neglect of spatial correlations. This
constraint, indeed, introduces 
some limitations which are particularly relevant in low dimensionality,
and in particular it
makes it impossible to treat phases with a definite spatial ordering such as $d$-wave superconductivity, $d$-density waves, stripes, dimerized states.

A few schemes have been proposed to overcome these limitations, re-introducing short-range correlations by replacing the single impurity model with a cluster-impurity, which contains $N_c$ sites in a given spatial arrangement.\cite{dca,cdmft,clustervari}
In this manuscript we compare two alternative schemes that represent somehow opposite perspectives in their ability to describe two-dimensional correlated models and $d$-wave superconductivity.
The Dynamical Cluster Approximation (DCA)\cite{dca} is based on a momentum-space perspective and it replaces the single momentum-independent self-energy of DMFT with the set of self-energies associated to the lattice momenta of an $N_c$-site cluster. For the (cluster) impurity model, this approach requires periodic boundary conditions on the cluster.
The other approach we consider, the Cellular Dynamical Mean-Field Theory (CDMFT), assumes instead a real space perspective, and it  generalizes more directly the mean-field spirit of DMFT\cite{cdmft}. In this scheme, a given cluster is chosen, and a ``local" theory for the cluster degrees of freedom is obtained through the cavity method, replacing the effect of the rest of the lattice with a self-consistent effective bath. The basic approximation is to assume that the dynamical field experienced by the cluster is Gaussian.

The properties of the two methods have been compared in several papers\cite{bk} which focused mainly on the asymptotic behavior for large clusters, and the two methods have been used to study many properties of the two-dimensional Hubbard model like, notably, $d$-wave superconductivity\cite{dwaveclusters}.
It must be underlined that, despite the simplifications introduced by the cluster methods with respect to the full lattice problem, the cluster-impurity model remains a non-trivial many-body problem, that still requires, in practice, a numerical ``solver'' in order to achieve the Green's functions. Among the most popular impurity solvers we remind various Quantum Monte Carlo methods (Hirsch-Fye determinantal method\cite{hirsch-fye}, and the more recently introduced Continuous-Time Quantum Monte Carlo\cite{ctqmc}), the exact-diagonalization approach\cite{ed}, the numerical renormalization group\cite{nrg}.

As a matter of fact, present ``state of the art'' calculations using accurate numerical solvers are limited to fairly small clusters\cite{plaquette,dwaveclusters}, or, if the cluster size is increased, to relatively small coupling and/or finite temperature\cite{largerclusters,tilted_DCA}, which may not be representative of the strong-repulsion regime. Such small sizes hardly allow for an accurate size scaling to describe the thermodynamic limit. It is therefore desirable to study the performance of the different cluster methods as a function of the cluster size within some approach that allows for a solution at arbitrary values of the size. This point of view has been taken in Ref.~\onlinecite{bk}, where a one-dimensional exactly solvable model has been studied using both CDMFT and DCA, allowing the comparison between the two methods. 

In this work we apply a similar strategy to treat a model which describes the essential physical ingredients of the cuprates, namely their two-dimensional character, the effects of strong correlations, and, most importantly, the presence of $d$-wave superconductivity. This model is the t-J model treated at a saddle point level, following, e.g., Ref.~\onlinecite{kotliarliu}. While this model is clearly an approximation of the full two-dimensional t-J model (which lacks an exact solution), we will consider it as our ``starting model''. 
In this way we will have an exactly solvable model, containing all the main ingredients of cuprates, that we can also solve using DCA and CDMFT approaches for any finite size of the clusters. 
This will allow us on one hand to study the convergence of the methods in the limit of large cluster-size, but on the other hand it will lead to a benchmark of the methods for small and intermediate clusters, such as those available in present numerical calculation and in those that can be expected in a few years.
Of particular interest, in this light, are the ``geometrical'' aspects of the different approaches. When a phase with a given spatial structure is present in a finite cluster, we can expect different behaviors according to the way in which the ordered phase fits in the chosen cluster. This will also depend on the boundary conditions, and will mark the difference between DCA and CDMFT.

The manuscript is organized as follows:
In Sec. II we present the reference model, i.e., the saddle-point t-J model. In Section. III we introduce DCA and CDMFT, and their application to our model. Sec. IV presents our results, and Sec. V contains concluding remarks.

\section{model and method}

\subsection{The saddle-point t-J model}
In this section we briefly review the derivation of the saddle-point t-J model 
in order to fix the notations and the main concepts.
Even if we will not attempt to solve it beyond saddle point, the 
starting point of our analysis is the two-dimensional t-J model
\begin{eqnarray}
 H &=&  P \left[ -\, t \sum_{\langle i,j\rangle} \left(f^\dagger_{i \sigma}
 f_{j \sigma} + \hc\right)  \,-\,
 \mu_0 \sum f^\dagger_{i\sigma} f_{i\sigma} \right. \nn\\
 & & \left. +\; J \sum_{\langle i,j\rangle} \left( {\mathbf{S}}_i \cdot {\mathbf{S}}_j -
 {\ts \frac{1}{4} } n_i n_j
 \right)  \right] P,
\end{eqnarray}
where ${\mathbf{S}}_i = \frac{1}{2} f^\dagger_{i\alpha}{\bf{\sigma}}_{\alpha\beta} f_{i\beta}$ is the local spin operator,
$n_i=\sum_\alpha f^\dagger_{i\alpha} f_{i\alpha}$ is the local
electron density and $P= \prod_i(1-n_{i\uparrow}n_{i\downarrow})$
is a projection operator which restricts the fermionic Hilbert
space to the low-energy subspace of empty and singly-occupied
sites; the super-exchange antiferromagnetic coupling $J$ is given
by $4t^2/U$.

We introduce slave boson fields $b_i$ in order to keep track of the empty
sites (holes): this representation allows, in fact, to replace the
constraint of zero double occupancy with the following equality constraint,
\begin{equation}
 \sum_\sigma f^\dagger_{i\sigma} f_{i\sigma} +
 b_i^\dagger b_i \;=\; 1, \label{eq:constr}
\end{equation}
where $b_i^\dagger b_i$ acquires the meaning of a local density of holes. Enforcing
this constraint by means of a local Lagrange multiplier
$\lambda_i$, we obtain:
\begin{eqnarray}
 H_\mathrm{sb} &=& -\, t \sum_{\langle i,j\rangle} \left(f^\dagger_{i \sigma}
 f_{j \sigma}\,  b_j^\dagger b_i  + \hc\right) \,-\,
 \mu_0 \sum f^\dagger_{i\sigma} f_{i\sigma} \nn\\
 & & +\, J \sum_{\langle i,j\rangle} \left( {\mathbf{S}}_i \cdot {\mathbf{S}}_j -
 {\ts \frac{1}{4} } (1 - b_i^\dagger b_i)(1 - b_j^\dagger b_j) \right)\nn\\
 & & + \sum \lambda_i \left( {\ts \sum_\sigma f^\dagger_{i\sigma} f_{i\sigma} +
 b_i^\dagger b_i - 1} \right).
\end{eqnarray}

To obtain an exactly solvable model we decouple the exchange interaction ${\mathbf{S}}_i \cdot {\mathbf{S}}_j$ 
introducing three sets of Hubbard-Stratonovich fields, which allow
us to treat on the same footing both the particle-hole and
particle-particle channels: the reason for this
approach is given by the requirement that the $SU(2)$
particle-hole symmetry at half-filling is being preserved. 

A static mean-field approximation is then achieved by replacing
the auxiliary fields and the Lagrange multiplier with their
saddle-point values. Setting $\langle {\mathbf{S}}_i
\rangle=0$ and neglecting the 4--boson hole-hole interaction,
which is $\ord(x^2)$ near half-filling ($x=\langle b_i^\dagger b_i
\rangle$ is the hole doping), the slave-boson mean-field
Hamiltonian reads
\begin{eqnarray}
 H^{\mathrm{MF}}_\mathrm{sb} &=& -\,t\sum_{\langle i,j\rangle} (f^\dagger_{i\sigma}
 f_{j\sigma} b_j^\dagger b_i  + \hc) +\nn\\
& &-\mu_f \sum f^\dagger_{i\sigma} f_{i\sigma} \,-\,
 \mu_b \sum b^\dagger_i b_i +\nn\\
  & & -\, \sum_{\langle i,j\rangle} (\chi_{ij}f^\dagger_{i\sigma}
 f_{j\sigma} + \hc)+\nn\\
  & & +\, \sum_{\langle i,j\rangle} \left( \Delta_{ij}
 (f^\dagger_{i\uparrow}f^\dagger_{j\downarrow}-f^\dagger_{i\downarrow}f^\dagger_{j\uparrow})
 + \hc \right), \label{eq:H_latt_b}
\end{eqnarray}
with the particle-hole and particle-particle amplitudes given by
\begin{eqnarray}
 \chi_{ij} &=& {\ts \frac{3}{8} } J\langle f^\dagger_{j\sigma}
 f_{i\sigma}\rangle,\label{eq:self1}\\
 \Delta_{ij} &=& {\ts \frac{3}{8} } J\langle f_{i\uparrow}f_{j\downarrow} -
 f_{i\downarrow}f_{j\uparrow}\rangle.\label{eq:self2}
\end{eqnarray}

The last step of our approximation consists in decoupling the
kinetic term of \ceq{eq:H_latt_b}, and this leads to an effective
hopping amplitude $\teff= t\langle b_j^\dagger b_i \rangle$ for
the fermionic degrees of freedom: assuming full boson condensation
at $T=0$, we can set $\langle b_j^\dagger b_i \rangle = {|\langle
b_i
\rangle |}^2 = x$ and consider, at last, $t_{\mathrm{eff}} = xt$.
At small doping the effects of strong correlations are thus
summarized in the renormalization of the free-fermion hopping
term, which leads to a strong suppression of the kinetic energy,
$\ord (xt)$, and to a relative enhancement of the
super-exchange energy, $\ord (J)$, i.e.\ finite, as $x\ra
0$.

In finding a self-consistent solution of the Hamiltonian
\eq{eq:H_latt_b}, it has been shown~\cite{Affleck:88} that while at half-filling there
are an infinite number of degenerate ground-states, connected together by
the $SU(2)$ rotations of the particle-hole symmetry, as soon as
doping breaks the $SU(2)$ invariance the lowest-energy state
is found to be~\cite{kotliarliu} the \dw solution $\chi_{\hat{x}}
=
\chi_{\hat{y}} = \chi$, $\Delta_{\hat{x}} = -
\Delta_{\hat{y}}=\Delta$.
We will therefore consider this kind of solution throughout our analysis.

Writing the fermionic part of $H^{\mathrm{MF}}_\mathrm{sb}$ in
Fourier space, we obtain
\begin{eqnarray}
 H_f &=& \sum_{\kk}(\epsilon_\kk-\mu) f^\dagger_{\kk \sigma} f_{\kk \sigma} \,+ \nn \\
 & & +\, \Delta_{\kk}
 (f^\dagger_{\kk\uparrow}f^\dagger_{-\kk\downarrow}+f_{-\kk\downarrow}f_{\kk\uparrow}),
 \label{eq:H_f}\\
 \epsilon_\kk &=& -2( t_{\mathrm{eff}} + \chi)(\cos k_x +\cos k_y),\\
 \Delta_\kk &=& 2\Delta (\cos k_x -\cos k_y),
\end{eqnarray}
where the parameters $\chi$ and $\Delta$, representing
respectively a renormalization of the hopping and a \dw order
parameter, are determined by the self-consistency equations
\eqs{eq:self1}{eq:self2}. 
These in Fourier space read
\begin{eqnarray}
 \chi &=& \frac{3}{8} J \int\! \frac{d^2 k}{{(2\pi)}^2}\,
 \langle f_\kk^\dagger f_\kk \rangle (\cos k_x + \cos k_y) \label{eq:self1latt} \\
  &=& \frac{3}{8} J \int\! \frac{d^2 k}{{(2\pi)}^2}\,
  \frac{(\mu -\epsilon_\kk)}{2 E_\kk}\tanh\frac{\beta E_\kk}{2}\, (\cos k_x + \cos k_y),
  \nn \\
 \Delta &=& \frac{3}{8} J \int\! \frac{d^2 k}{{(2\pi)}^2}\,
 \langle f_{\kk\uparrow} f_{-\kk\downarrow} \rangle (\cos k_x - \cos k_y) \label{eq:self2latt} \\
  &=& \frac{3}{8} J \int\! \frac{d^2 k}{{(2\pi)}^2}\,
  \frac{\Delta_\kk}{2 E_\kk}\tanh\frac{\beta E_\kk}{2}\, (\cos k_x - \cos k_y),
  \nn
\end{eqnarray}
where $E_\kk = \sqrt{{(\epsilon_\kk -\mu)}^2+\Delta_\kk^2}$ are
the eigenvalues of the Hamiltonian \eq{eq:H_f}. The fermionic
chemical potential $\mu$ is instead determined by the number equation
\begin{equation}
 1-x \;=\; 1 - \int\! \frac{d^2 k}{{(2\pi)}^2}\,
 \frac{(\epsilon_\kk - \mu)}{E_\kk}\tanh\frac{\beta E_\kk}{2}, \label{eq:mulatt}
\end{equation}
obtained by imposing the fermion density to be $(1-x)$.

\subsection{Cluster Approximations}
We are now in the position to compare the exact solution of the saddle-point model (\ref{eq:H_latt_b}) 
with the approximate cluster solutions.
Within cluster DMFT methods an effective action for the cluster degrees of freedom is defined as
\begin{eqnarray}
S_\mathrm{eff}&=&\int_{-\beta}^{\beta}d\tau \sum_{\mu\nu\sigma} 
c^{\dagger}_{\mu\sigma}(\tau)
{{\cal{G}}^{-1}(\tau-\tau^{\prime})}_{\mu\nu}c_{\nu\sigma}
(\tau^{\prime})+\nonumber\\
&+& \int_{-\beta}^{\beta} d\tau \sum_{\mu=1}^{N_c} U n_{\mu\uparrow}(\tau) 
n_{\mu\downarrow}(\tau),
\label{seff}
\end{eqnarray}
where ${{\cal{G}}^{-1}}$ is a dynamical ``Weiss'' field. By computing the cluster Green's function  $G_{\mu\nu\sigma}(\tau) = 
-\langle T c_{\mu\sigma}(\tau) c^{\dagger}_{\nu\sigma} \rangle$,
the cluster self-energy is obtained as
\begin{equation}
\Sigma^{\mu \nu}_{c}(i\omega_{n})=\mathcal{G}_{\mu \nu}^{-1}(i\omega_{n})-
 G_{\mu \nu}^{-1}(i\omega_{n}).
\label{selfenergy}
\end{equation}
The two methods differ in the way the new Weiss field is obtained through
the knowledge of the cluster self-energy $\Sigma^{\mu \nu}_{c}(i\omega_{n})$.

Within CDMFT the ``local'' Green's function for the cluster is calculated as
\begin{equation}
 G_{loc}^{-1}(i\omega_{n})= \int_{\pi/N_{c}}^{\pi/N_{c}}
\frac{1}{i\omega_{n}+ \mu- {t}_{k}- \Sigma_{c}(i\omega_{n})}
 \frac{dk}{2\pi/N_{c}} ,
\label{Gloc}
\end{equation}
where the momentum-integral extends over the reduced Brillouin zone 
associated to the $N_c$-site cluster, $t_{k}$ is the Fourier transform
of the cluster hopping term.
$G_{loc}(i\omega_n)^{\mu\nu}$ is then used to obtain a new
Weiss field
\begin{equation}
(\mathcal{G}_0^{new})_{\mu \nu}^{-1}(i\omega_{n})= \,\Sigma^{\mu
  \nu}_{c}(i\omega_{n})+ G_{ loc }^{-1 \, \mu\nu }(i\omega_{n}),
\label{neweiss}
\end{equation}
which determines the new effective action (\ref{seff})
from which a new $G_{\mu\nu}(i\omega_n)$ can be obtained: the 
procedure is then iterated until convergence.
We stress that this method does not impose lattice translational invariance.

The spirit of DCA is instead to generalize the momentum-independence of the
self-energy,
characteristic of the single-site DMFT, to a small cluster. Thus one defines
a coarse-grained self-energy for every reciprocal lattice momentum
$\kk_c$
associated to the cluster at hand.
The analogue of (\ref{Gloc}), which expresses the lattice
Green's function in terms of the cluster self-energy, is given by
\begin{equation}
 G(\kk_c + \KK,\iom) = \frac{1}%
 {\iom +\mu - t(\kk_c+\KK) - \Sigma^c(\kk_c,\iom)}\,, \label{eq:green_latt_DCA}
\end{equation}
while the self-consistency relation between the Weiss field and
the cluster self-energy now reads
\begin{equation}
\G_0^{-1}(\kk_c,\iom) = \left[\frac{N_c}{N} \sum_\KK G(\kk_c + \KK,\iom)\right]^{-1} +
\Sigma^c(\kk_c,\iom). \label{eq:self_DCA}
\end{equation}
In these expressions, all the cluster quantities appear as
functions of the cluster momenta $\kk_c$, and the $\KK$
integration over the reduced Brillouin zone  is nothing but a coarse-graining of the
lattice Green's function around these momenta.

A crucial observation is that the diagonal nature in momentum space of the
DCA equations requires that the cluster part of our effective action has periodic
boundary conditions. As we will discuss, this may represent a severe constraint, especially for
small cluster sizes.

Even if DCA is naturally defined in momentum space, it is
useful to recover a real-space formulation also for this cluster
approach, in order to have a unified formalism which makes easier
the comparison between the two methods. Performing a Fourier
transform upon the cluster momenta, all the cluster quantities
$\mathcal{Q}(\kk_c)$ become cyclic matrices in the real-space cluster
indexes, i.e., with the matrix elements $\mathcal{Q}_{ij}$ depending only
on $(i-j)\!\mod\! L_c$: this means that translational invariance
is preserved within the cluster, which thus must have periodic
boundary conditions, as we mentioned before. The real space
formulation of the self-consistency equation \eq{eq:self_DCA} is
then given by
\begin{eqnarray}
 & & \hat{\G}_0^{-1}(\iom) \,=\, \hat{\Sigma}^c(\iom) \, + \ph{\int} \label{eq:self_DCA_real} \\ 
 & & + \, \left[\frac{N_c}{N} \sum_\KK
 \left[ (\iom +\mu)\hat{1} - \hat{t}_\DCA(\KK) - \hat{\Sigma}^c(\iom) \right]^{-1} \right]^{-1}, \nn
\end{eqnarray}
where ${[\hat{t}_\DCA(\KK)]}_{ij}=\frac{1}{N_c}\sum_{\kk_c}
e^{i\kk_c \cdot (\ee_i - \ee_j)} t(\kk_c+\KK) =
e^{-i\KK\cdot(\ee_i-\ee_j)} t_{ij}(\KK)$ differs from the bare
$\hat{t}(\KK)$ used in CDMFT in order to satisfy the cyclicity
condition, i.e., the translational invariance.
As in CDMFT, the solution requires an iterative solution of a
cluster-impurity
model determined self-consistently.

We now detail the implementation of the two approaches for the
mean-field two-dimensional t-J model.

\subsection{DCA for the saddle-point t-J model}

In our analysis of the \tJ model, the DCA cluster self-energy
$\Sigma(\kk_c)$ consists of a normal term $\chi(\kk_c)$ and of an
anomalous term $\Delta(\kk_c)$, which are respectively the Fourier
transform of $\chi_{ij}$ and $\Delta_{ij}$ within the DCA
cluster; explicitly, they are given by
\begin{eqnarray}
 \chi(\kk_c) &=&
 -2\chi_{cl}^{\DCA} (\cos{k_c}_x +\cos{k_c}_y),\\
 \Delta(\kk_c) &=&
 2\Delta_{cl}^{\DCA} (\cos{k_c}_x -\cos{k_c}_y).
\end{eqnarray}
Since our starting model is a mean-field model, these quantities are $\omega$--independent; however,
this deficiency is compensated by the possibility to solve    
exactly both the lattice and the cluster problems, even for large
values of $N_c$.

The cluster parameters $\chi_{cl}^{\DCA}$ and $\Delta_{cl}^{\DCA}$
are determined by the DCA self-consistency equations
\begin{eqnarray}
 \chi_{cl}^{\DCA} &=& \frac{3}{8} J \int\! \frac{d^2 k}{{(2\pi)}^2}\,
 \langle f_\kk^\dagger f_\kk \rangle (\cos{k_c}_x + \cos{k_c}_y) \nn \\
  &=& \frac{3}{8} J \sum_{\kk_c}\int\! \frac{d^2 K}{{(2\pi)}^2}\,
 \frac{(\mu -\epsilon_{\KK,\kk_c})}{2 E_{\KK,\kk_c}} \times \label{eq:self1DCA} \\
  &\times& \tanh\frac{\beta E_{\KK,\kk_c}}{2}\,
 (\cos{k_c}_x + \cos{k_c}_y),\nn
\end{eqnarray}
\begin{eqnarray}
 \Delta_{cl}^{\DCA} &=& \frac{3}{8} J \int\! \frac{d^2 k}{{(2\pi)}^2}\,
 \langle f_{\kk\uparrow} f_{-\kk\downarrow} \rangle (\cos{k_c}_x - \cos{k_c}_y) \nn \\
  &=& \frac{3}{8} J \sum_{\kk_c}\int\! \frac{d^2 K}{{(2\pi)}^2}\,
 \frac{\Delta_{\kk_c}}{2 E_{\KK,\kk_c}}\times \label{eq:self2DCA} \\
  &\times& \tanh\frac{\beta E_{\KK,\kk_c}}{2}\,
 (\cos{k_c}_x - \cos{k_c}_y),\nn
\end{eqnarray}
where $\epsilon_{\KK,\kk_c} = -2\teff \left[\cos({k_c}_x + K_x) +
\cos({k_c}_y + K_y)\right] + \chi(\kk_c)$ and $E_{\KK,\kk_c} =
\sqrt{{(\epsilon_{\KK,\kk_c}-\mu)}^2+\Delta_{\kk_c}^2}$. In the second row of
\ceqs{eq:self1DCA}{eq:self2DCA} the integration over the entire Brillouin zone
is divided into a sum over the cluster momenta and an integration
over the reduced Brillouin zone, $\int\! \frac{d^2 k}{{(2\pi)}^2}
\equiv
\frac{1}{L_c^2}\sum_{\kk_c} \int\! \frac{d^2 K}{\left(\frac{2\pi}{L_c}\right)^2}$.

It is important to note that these cluster quantities, used in the
definition of the cluster self-energy, do not have an immediate
physical meaning and thus they cannot be directly compared to the
corresponding lattice quantities of
\ceqs{eq:self1latt}{eq:self2latt}. The physically relevant quantities are
instead given by
\begin{eqnarray}
 \chi_{latt}^{\DCA} &=& \frac{3}{8} J \int\! \frac{d^2 k}{{(2\pi)}^2}\,
 \langle f_\kk^\dagger f_\kk \rangle (\cos{k}_x + \cos{k}_y), \\
 \Delta_{latt}^{\DCA} &=& \frac{3}{8} J \int\! \frac{d^2 k}{{(2\pi)}^2}\,
 \langle f_{\kk\uparrow} f_{-\kk\downarrow} \rangle (\cos{k}_x - \cos{k}_y), \ph{oo}
\end{eqnarray}
where the expectation values must be evaluated using the
self-consistent parameters $\chi_{cl}^{\DCA}$ and
$\Delta_{cl}^{\DCA}$, as in
\ceqs{eq:self1DCA}{eq:self2DCA}.

Finally, the DCA analogue of \ceq{eq:mulatt} gives the
self-consistency equation for the chemical potential:
\begin{eqnarray}
 1-x &=& 2 \int\! \frac{d^2 k}{{(2\pi)}^2}\,
 \langle f_\kk^\dagger f_\kk \rangle \\
  &=& 1 - \sum_{\kk_c}\int\! \frac{d^2 K}{{(2\pi)}^2}\,
 \frac{(\epsilon_{\KK,\kk_c} -\mu)}{E_{\KK,\kk_c}}
 \tanh\frac{\beta E_{\KK,\kk_c}}{2}. \nn
\end{eqnarray}

\subsection{CDMFT for the saddle-point t-J model}
\label{subsec:CDMFT}

Considering a square cluster $\cl$ with $N_c=L_c\times L_c$ sites,
we denote by $i\equiv i(i_x,i_y)=i_x+(i_y-1)L_c$ the cluster site
with coordinates $(i_x,i_y)$, where $i_x,i_y=1,\ldots,L_c$. The
local cluster Hamiltonian is then obtained from the lattice one by
restricting all the site-index sums to the cluster sites:
\begin{eqnarray}
 H_c &=& \sum_{i,j\,\in\, \cl} (t_{ij}+\chi_{ij}^c)
 f^\dagger_{i\sigma} f_{j\sigma} \,-\, \mu\sum_{i\,\in\,\cl}
 f^\dagger_{i\sigma} f_{i\sigma} \,+\nn\\ &+&\sum_{i,j\,\in\, \cl}
 (\Delta_{ij}^c f^\dagger_{i\uparrow}f^\dagger_{j\downarrow} + \hc),\ph{ooo}
\end{eqnarray}
where $\hat{\chi}^c$ and $\hat{t}$ are hermitian matrices and
$\hat{\Delta}^c$ a symmetric matrix. Explicitly, their expressions
read
\begin{eqnarray}
 t_{ij} &=& -\teff \sum_{\hat{\eta}}\left( \delta_{j,i+\hat{\eta}} + \delta_{j,i-\hat{\eta}}\right),\\
 \chi_{ij}^c &=& - \sum_{\hat{\eta}}\left( \chi_{i,\hat{\eta}}\delta_{j,i+\hat{\eta}} + \chi^*_{j,\hat{\eta}}\delta_{j,i-\hat{\eta}} \right), 
 \ph{ooo}\label{eq:chimatrix}\\
 \Delta_{ij}^c &=& \sum_{\hat{\eta}}\left(\Delta_{i,\hat{\eta}}\delta_{j,i+\hat{\eta}} +
 \Delta_{j,\hat{\eta}}\delta_{j,i-\hat{\eta}}\right)
 ,\label{eq:delmatrix}
\end{eqnarray}
where $\hat{\eta} = \hat{x},\hat{y}$ is a lattice displacement in the x or y direction, $\delta_{j,i+\hat{x}}=\delta_{j,i+1}(1-\sum_n\delta_{i,nL_c})$,
$\;\delta_{j,i-\hat{x}}=\delta_{j,i-1}(1-\sum_n\delta_{i,nL_c+1})$
and $\delta_{j,i\pm\hat{y}}=\delta_{j,i\pm L_c}$. The
self-consistent parameters $\chi_{i,\hat{\eta}}$ and $\Delta_{i,\hat{\eta}}$ are formally given
by \ceqs{eq:self1}{eq:self2}, where the
expectation values must be evaluated using the cluster propagator
$\hat{D}_c$, defined below.

In order to write this propagator in a compact form, we first
introduce the Nambu spinor $\Psi^\dagger \equiv
(f_{1\uparrow}^\dagger, \ldots, f_{L_c^2\uparrow}^\dagger,
f_{1\downarrow}, \ldots, f_{L_c^2\downarrow})$, which contain all
the $2 L_c^2$ fermionic degrees of freedom within the cluster.
With this notation,
\begin{equation}
 \hat{D}_c(\tau) = -\langle T\Psi(\tau)\Psi^\dagger(0)\rangle = \left(
 \begin{array}{cc}
  \hat{G}_\uparrow(\tau) & \hat{F}^\dagger(-\tau) \\
  \hat{F}(\tau) & -\hat{G}^T_\downarrow(-\tau) \\
 \end{array}
 \right),
\end{equation}
where $G_{ij,\,\sigma}=-\langle
Tf_{i\sigma}(\tau)f_{j\sigma}^\dagger(0) \rangle$ and
$F_{ij}=-\langle
Tf_{i\downarrow}^\dagger(\tau)f_{j\uparrow}^\dagger(0) \rangle$
are, respectively, the normal and anomalous Green's functions. We
can then express $\hat{D}_c$ in terms of the cluster Hamiltonian
parameters,
\begin{equation}
 \hat{D}_c(i\omega_n) = \int\! \frac{d^2 K}{\left(\frac{2\pi}{L_c}\right)^2}
 \, {\left[ i\omega_n\hat{1} - \hat{h}(\KK)\right]}^{-1},
\end{equation}
\begin{equation}
 \hat{h}(\KK) = \left(
 \begin{array}{cc}
  \hat{t}(\KK)+\hat{\chi}^c-\hat{\mu} &  \hat{\Delta}^c \\
  {{(\hat{\Delta}^c)}^*}^{\phantom{\int}} &  -{\left[\hat{t}(-\KK)+\hat{\chi}^c-\hat{\mu}\right]}^*\\
 \end{array}
 \right), \ph{oo}
\end{equation}
where $\hat{\mu}$ is the chemical potential times the unitary matrix, $\hat{t}(\KK)$ is the Fourier transform of the super-lattice
hopping matrix $\hat{t}_{\mathbf{R},\mathbf{R'}}$, which, for
$|\mathbf{R}-\mathbf{R'}|=L_c$, connects the boundary sites of
neighboring clusters:
\begin{eqnarray}
 t(\KK)_{ij} &=& t_{ij} \,-\, \teff
 \sum_{n=1}^{L_c} \left[\exp(iK_xL_c) \delta_{i,nL_c}\delta_{j,(n-1)L_c+1}
 \right. \nn \\
 & &  +\, \exp(-iK_xL_c) \delta_{j,nL_c}\delta_{i,(n-1)L_c+1} \,+ \nn \\
 & &  +\, \exp(iK_yL_c) \delta_{j,n}\delta_{i,L_c(L_c-1)+n}   \,+ \nn \\
 & & \left. +\,
 \exp(-iK_yL_c) \delta_{i,n}\delta_{j,L_c(L_c-1)+n} \right].
\end{eqnarray}
Introducing the unitary matrix $\hat{U}(\KK)$ which diagonalizes $\hat{h}(\KK)$,
\begin{equation}
 \left[\hat{U}(\KK) \hat{h}(\KK) \hat{U}^\dagger(\KK)\right]_{\mu\nu} =
 \delta_{\mu,\nu}\lambda^\nu(\KK),
\end{equation}
we can write the cluster propagator as (the dependence of the matrices on $\KK$ is omitted 
to lighten the notation)
\begin{equation}
 \left[\hat{D}_c(i\omega_n)\right]_{IJ}= \int\! \frac{d^2 K}{\left(\frac{2\pi}{L_c}\right)^2}\,
 \sum_{\nu=1}^{2L_c^2} {\hat{U}}^*_{\nu I}\,\hat{U}_{\nu J}\,
 \frac{1}{i\omega_n-\lambda^\nu},
\end{equation}
where the index $I\equiv(\sigma,i)$ denotes the $(\sigma-1)L_c^2
+i$ component of a Nambu spinor ($\sigma =1,2$).

It is now easy to get the self-consistency relations for the
cluster parameters:
\begin{eqnarray}
 \chi_{i,\hat{x}} &=& \frac{3}{8} J \int\! \frac{d^2 K}{\left(\frac{2\pi}{L_c}\right)^2}\,
 \sum_{\nu=1}^{2L_c^2}\left[{\hat{U}}^*_{\nu,\,(1,i)}\,\hat{U}_{\nu,\,(1,i+1)}\right.
 \label{eq:chiCDMFT}\\
 & & \left. -\, {\hat{U}}^*_{\nu,\,(2,i+1)}\,\hat{U}_{\nu,\,(2,i)}\right]
 f\left(\beta \lambda^\nu\right), \qquad i\neq n L_c\nn\\
 \chi_{i,\hat{y}} &=& \frac{3}{8} J \int\! \frac{d^2 K}{\left(\frac{2\pi}{L_c}\right)^2}\,
 \sum_{\nu=1}^{2L_c^2} \left[{\hat{U}}^*_{\nu,\,(1,i)}\,\hat{U}_{\nu,\,(1,i+L_c)}\right.\\
 & & \left. -\, {\hat{U}}^*_{\nu,\,(2,i+L_c)}\,\hat{U}_{\nu,\,(2,i)}\right]
 f\left(\beta \lambda^\nu\right), \quad i\leq L_c(L_c-1)\nn
\end{eqnarray}
\begin{eqnarray}
 \Delta_{i,\hat{x}} &=& -\frac{3}{8} J \int\! \frac{d^2 K}{\left(\frac{2\pi}{L_c}\right)^2}\,
 \sum_{\nu=1}^{2L_c^2} \left[{\hat{U}}^*_{\nu,\,(1,i)}\,\hat{U}_{\nu,\,(2,i+1)}\right.\\
 & & \left. +\, {\hat{U}}^*_{\nu,\,(1,i+1)}\,\hat{U}_{\nu,\,(2,i)}\right]
 f\left(\beta \lambda^\nu\right), \qquad i\neq n L_c\nn\\
 \Delta_{i,\hat{y}} &=& -\frac{3}{8} J \int\! \frac{d^2 K}{\left(\frac{2\pi}{L_c}\right)^2}\,
 \sum_{\nu=1}^{2L_c^2} \left[{\hat{U}}^*_{\nu,\,(1,i)}\,\hat{U}_{\nu,\,(2,i+L_c)}\right.
 \label{eq:delCDMFT}\\
 & & \left. +\, {\hat{U}}^*_{\nu,\,(1,i+L_c)}\,\hat{U}_{\nu,\,(2,i)}\right]
 f\left(\beta \lambda^\nu\right), \quad i\leq L_c(L_c-1)\nn
\end{eqnarray}
where $f(x)={(e^x+1)}^{-1}$ is the Fermi function.

The determination of the chemical potential $\mu$ as a function of the fermion density
requires some more care. In fact, as
we have seen in \ceqs{eq:chiCDMFT}{eq:delCDMFT} for  $\chi$ and $\Delta$, 
the observables are generally site-dependent in CDMFT, and there is no
unique procedure to extract lattice observables
(translationally invariant) from cluster quantities. In general,
for a local observable $\ord_i$, we can estimate its lattice
counterpart with a weighted average
\begin{equation}
\ord_{latt} = \sum_{i\,\in\,\cl} w_i \ord_i,
\end{equation}
where $\sum_{i\,\in\,\cl} w_i = 1$, and we have decided to
investigate the two extreme cases of flat average, $w^{\flat}_i =
1/N_c$, and bulk value, $w^{\bulk}_i =
\delta_{i,b}$, the latter case corresponding to taking the value of the
observable just in the center of the cluster \cite{notacenter}, represented by the site $b$. From the
local fermion density
\begin{equation}
 \langle n_i \rangle \;=\; 1 + \int\! \frac{d^2 K}{\left(\frac{2\pi}{L_c}\right)^2}\,
 \sum_{\nu=1}^{2L_c^2}\left[ {\left|{\hat{U}}_{\nu,\,(1,i)}\right|}^2
 - {\left|{\hat{U}}_{\nu,\,(2,i)}\right|}^2 \right]
 f\left(\beta \lambda^\nu\right)
\end{equation}
we can therefore extract an average density $n^{\flat} = 1/L_c^2
\sum_i \langle n_i \rangle$ and a bulk density $n^{\bulk} =
\langle n_{b} \rangle$, and we can adjust the chemical potential
in order to satisfy either one of the two equations
\begin{equation}
n^{(\flat,\,\bulk)} = 1-x. \label{eq:flat_bulk_dens}
\end{equation}

The same argument would apply in extracting the lattice
self-energy parameters, to be compared with those of
\ceqs{eq:self1latt}{eq:self2latt}, from the corresponding cluster
quantities. However, in considering the flat average case, we
should note that $\chi_{ij}$ and $\Delta_{ij}$ are defined on
bonds, so that for each direction their total number is
$L_c(L_c-1)$ instead of $L_c^2$; the averages will thus be given
by
\begin{eqnarray}
 \chi^{\flat}_{\hat{x}} &=& \frac{1}{L_c(L_c-1)}\sum_{i\neq nL_c}\chi_{i,\hat{x}}\,,\\
 \chi^{\flat}_{\hat{y}} &=& \frac{1}{L_c(L_c-1)}\sum_{i\leq L_c(L_c-1)}\chi_{i,\hat{y}}\,,
\end{eqnarray}
\begin{eqnarray}
 \Delta^{\flat}_{\hat{x}} &=& \frac{1}{L_c(L_c-1)}\sum_{i\neq nL_c}\Delta_{i,\hat{x}}\,,\label{eq:flatDelta1}\\
 \Delta^{\flat}_{\hat{y}} &=& \frac{1}{L_c(L_c-1)}\sum_{i\leq L_c(L_c-1)}\Delta_{i,\hat{y}}\,.\label{eq:flatDelta2}
\end{eqnarray}
The bulk value estimate is instead straightforward and corresponds
to the values of the parameters on the innermost bonds of the cluster.

\section{results}

\subsection{Size dependence of observables}
We start our analysis of DCA and CDMFT for the saddle-point t-J model by
considering the behavior of three relevant observables as a function of the linear size of
the cluster for square lattices of $L_c \times L_c$ sites. Throughout this section, energy 
scales are expressed in units of $J/4$ and $J/t=0.4$.
\begin{figure}[h]
\includegraphics[width=8cm]{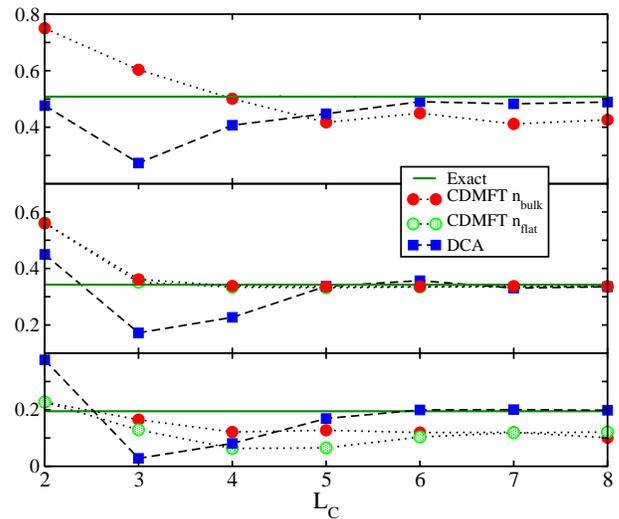}
\caption{(Color online). Normal parameter $\Delta$ as a function of the linear dimension $L_c$ for square clusters and different cluster methods. Red dots are CDMFT with bulk density, green dots CDMFT with average density, blue squares DCA. The thermodynamic limit is marked by the thick green line. From top to bottom $x=0, 0.1, 0.2$.
}\label{deltavslc}
\end{figure}
In Fig. \ref{deltavslc} we plot the superconducting order parameter for
different values of doping $x=0,0.1$ and $0.2$. As discussed above, for CDMFT one
has the alternative between ``bulk'' and ``average'' estimates. With respect to $\Delta$, we have found that the average estimate 
is much less sensitive to the size of the system than the bulk one, the latter being instead rather unreliable,
compared to the exact solution, up to large values of $L_c$.
We will thus present, in each panel, only average estimates of $\Delta$, while considering
both the average and the bulk estimates of the electron density.
The behavior of $\Delta$ as a function of the cluster size is completely non trivial and reveals important differences between the two methods. While DCA converges smoothly and faster from $L_c = 5$, it presents strong size-effects for smaller clusters: the $L_c=2$ cluster overestimates $\Delta$ at large doping, while $L_c=3$ and $4$ produce a strongly underestimated value of $\Delta$. This shows that momentum-space discretization is too strong to properly describe the spatial structure of the order parameter as long as the number of allowed momenta is small.
On the other hand, while for $L_c=2$ CDMFT overestimates $\Delta$ as well, the results obtained for small clusters are in general
more reliable, with relatively small deviations from the exact solution and, most important, a smoother dependence on $L_c$. 
However, this method converges more slowly to the thermodynamic limit, in particular for larger dopings, where $\Delta$
is systematically underestimated.
The comparison between different doping values shows indeed that CDMFT is quite inaccurate for $x=0.2$, a doping value at which the hopping processes become more relevant, according to $t_\mathrm{eff}=xt$, making the system itinerant and consequently better described in momentum space than in real space. The bulk estimate of the density is typically found to provide a better agreement with the exact solution.

\begin{figure}[h]
\includegraphics[width=8cm]{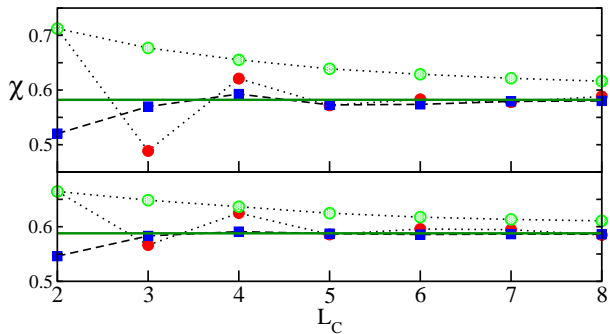}
\caption{(Color online) Normal parameter $\chi$ as a function of the linear dimension $L_c$ for square clusters and different cluster methods. 
Red dots are CDMFT bulk values of $\chi$, green dots CDMFT average values, blue squares DCA. The thermodynamic limit is marked by the thick green line.
From top to bottom $x=0.1, 0.2$.
}\label{chivslc}
\end{figure}

The same tendencies are present in the critical temperature (not shown), with a significantly enhanced overestimate of $T_c$ compared to $\Delta$ in the $L_c=2$ DCA cluster (we obtain $T_c =$ 1.53, 1.46 and 1.30 for $x = 0, 0.1, 0.2$, while the exact results are 0.767, 0.601 and 0.374). The systematic underestimate of $T_c$ for large doping in CDMFT is also reflected in a smaller critical doping at which superconductivity disappears ($x_c \sim 0.25$ even for large clusters, as opposed to the thermodynamic limit $x_c\simeq 0.35$).

We finally consider the normal parameter $\chi$. Since this parameter coincides with $\Delta$ at half-filling, due to the particle-hole symmetry, we focus only on $x=0.1$ and $0.2$. Here, in agreement with previous studies in a one dimensional model \cite{bk}, we find that in CDMFT the bulk estimate of the parameter is more accurate than the average one. Yet, CDMFT is less accurate than DCA for the cluster we studied, signaling that the exponential convergence of bulk estimates is established only for larger values of $L_c$. 

The different behavior between the superconducting and the normal parameters underlines that the accuracy of the different approaches, for small clusters, depends crucially on the quantity under consideration. In particular, the $d$-wave superconducting order parameter suffers stronger size effects due to its peculiar structure in real-space, and, when the number of cluster sites becomes small, it is better represented by the CDMFT solution. On the other hand, the normal parameter, which is more isotropic, is better described by DCA, which favors homogeneous states (in the sense of states without peculiar patterns).

\subsection{Doping dependence for small clusters}
In this section we focus on the smallest clusters $L_c=2,3,4,5$, where we analyze in more details the doping dependence of the observables. This study is of particular interest because only small sizes can be handled in a full numerical solution of the Hubbard or t-J models using CDMFT and DCA approaches. For the sake of definiteness, in CDMFT we use the flat average estimate for both $\Delta$ and $\chi$ and the bulk estimate for the electron density.
\begin{figure}[h]
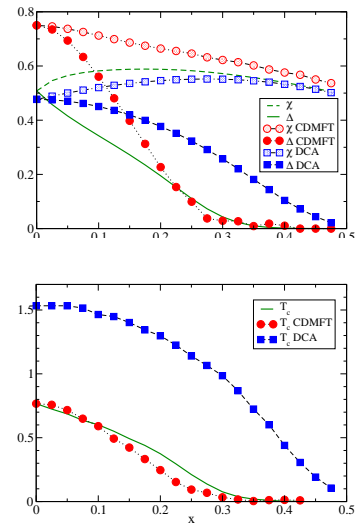

\includegraphics[width=4.5cm]{fig3a.eps}
\vskip 0.4cm
\includegraphics[width=4.5cm]{fig3b.eps}
\caption{(Color online) Doping dependence of $\Delta$ and $\chi$ (top) and of the critical temperature $T_c$ (bottom) as a function of doping for a $2\times 2$ cluster and different cluster methods. Filled symbols refer to $\Delta$, open symbols to $\chi$. Red dots are used for CDMFT, blue squares for DCA. The solid green line is the thermodynamic limit.
}\label{Lc2}
\end{figure}

\vspace{0.4cm}
\begin{figure}[h]
\includegraphics[width=4.5cm]{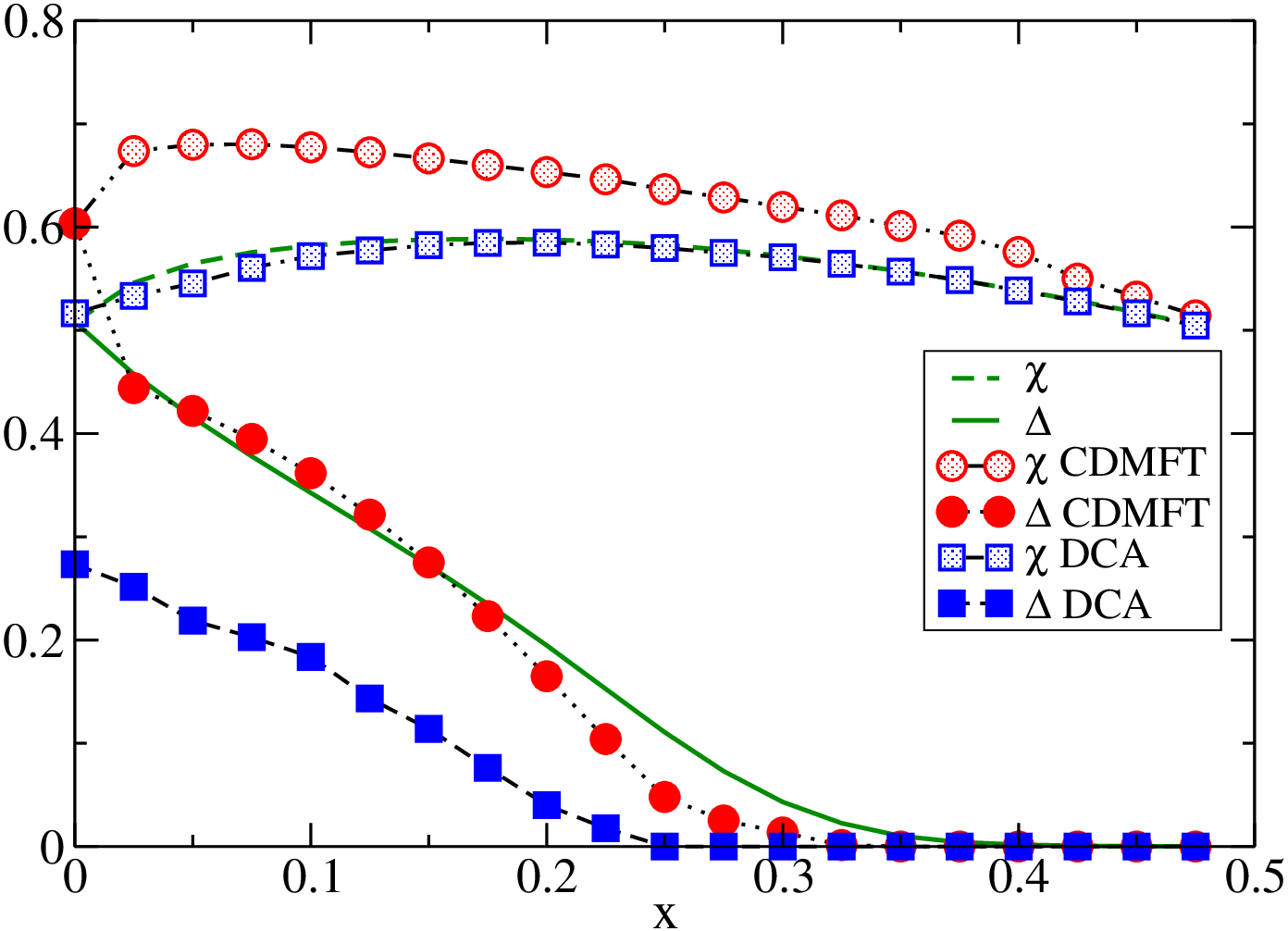}
\vskip 0.4cm
\includegraphics[width=4.5cm]{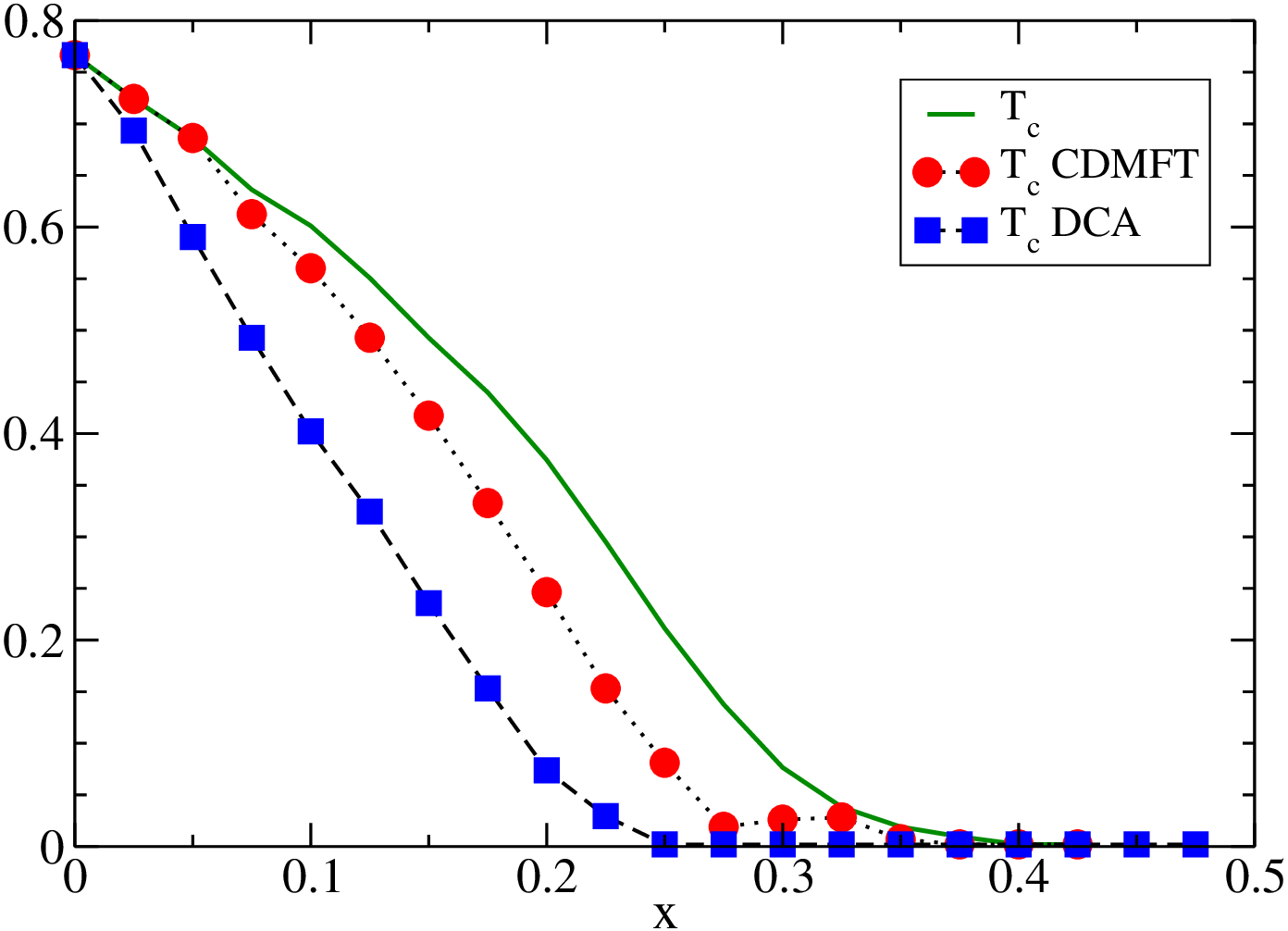}
\caption{(Color online) Same as Fig. \ref{Lc2} for $3\times 3$ cluster.}\label{Lc3}
\end{figure}

\vspace{0.4cm}
\begin{figure}[h]
\includegraphics[width=4.5cm]{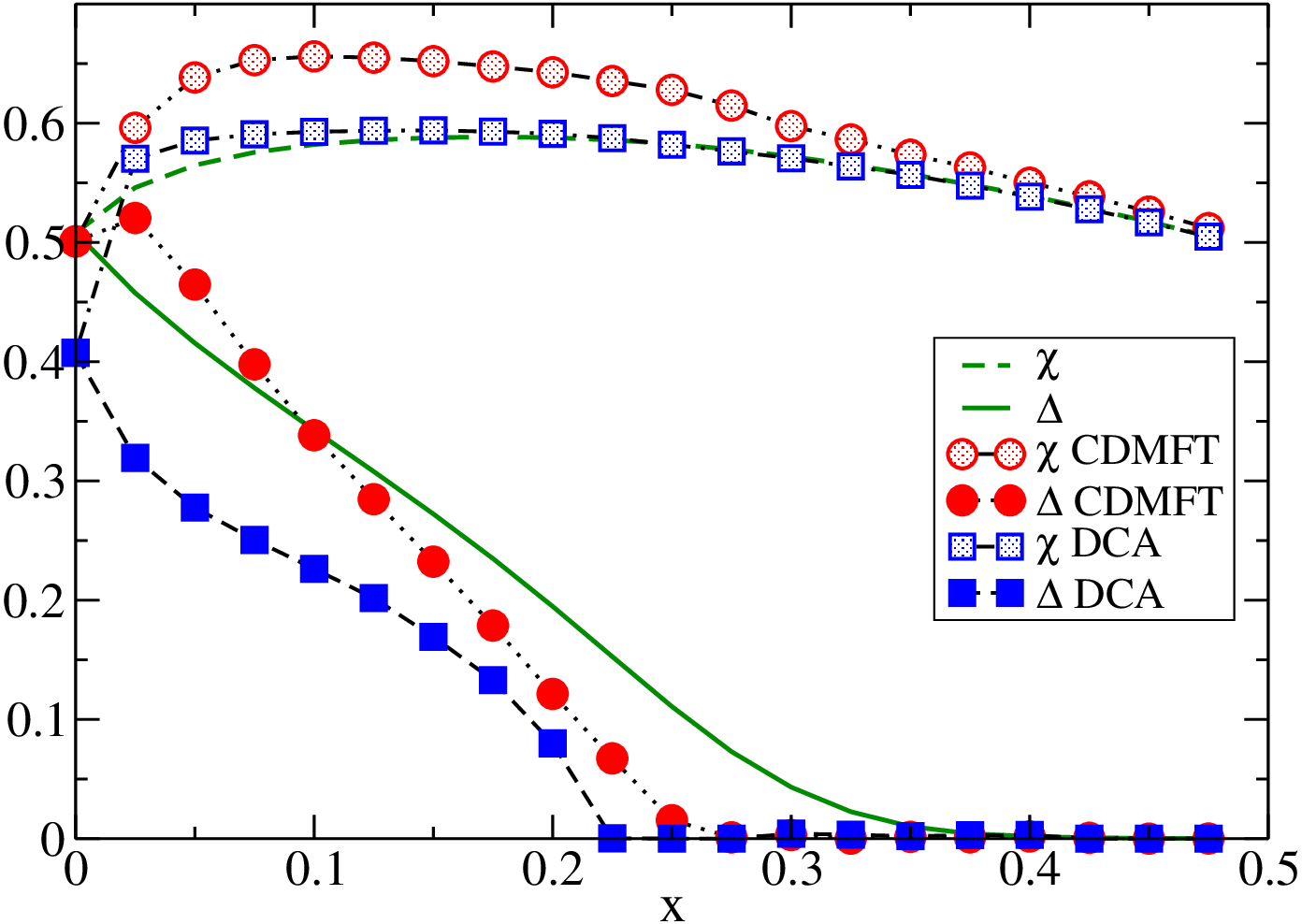}
\vskip 0.4cm
\includegraphics[width=4.5cm]{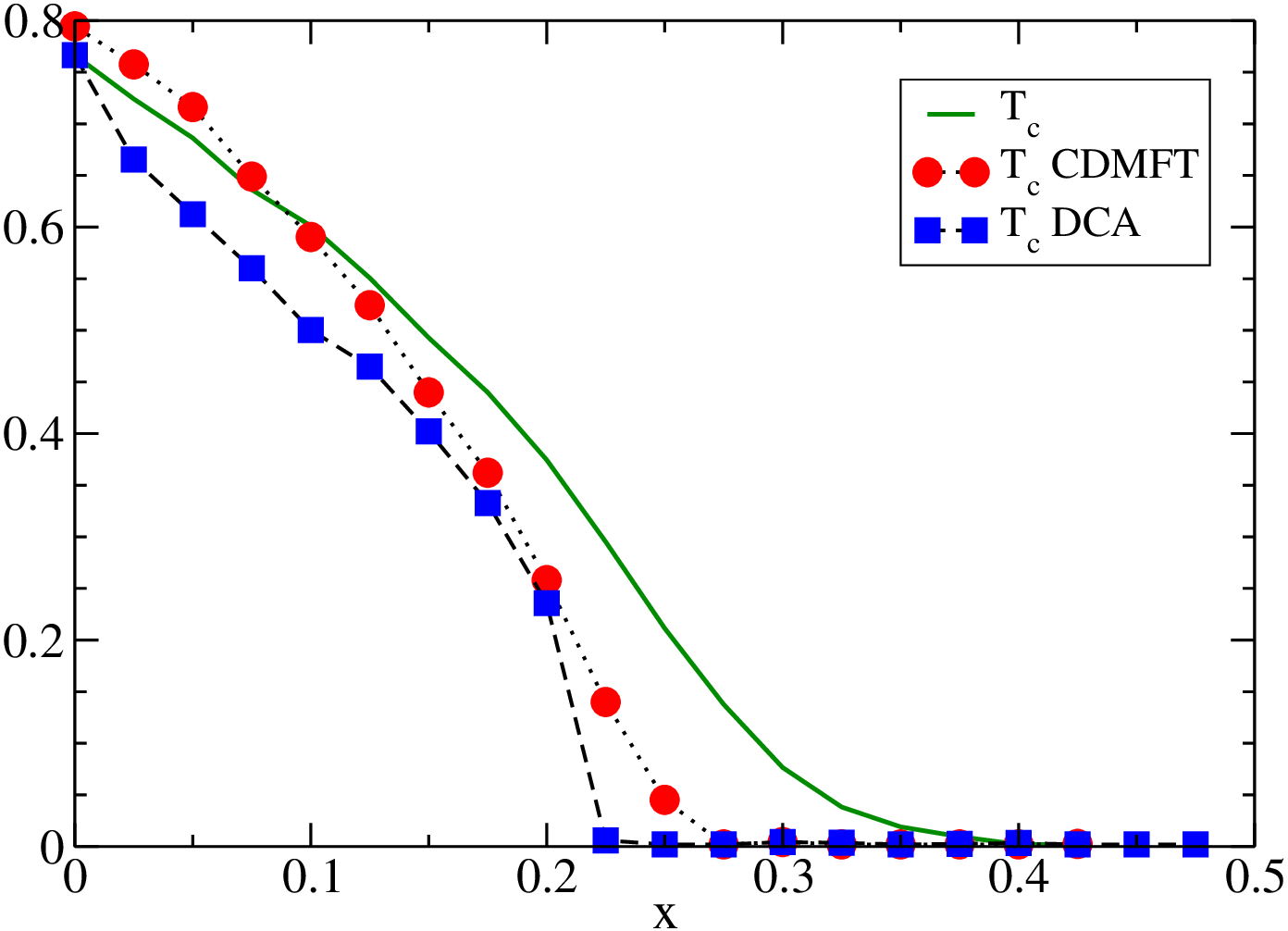}
\caption{(Color online) Same as Fig. \ref{Lc2} for $4\times 4$ cluster.
}\label{Lc4}
\end{figure}

\vspace{0.4cm}
\begin{figure}
\includegraphics[width=4.5cm]{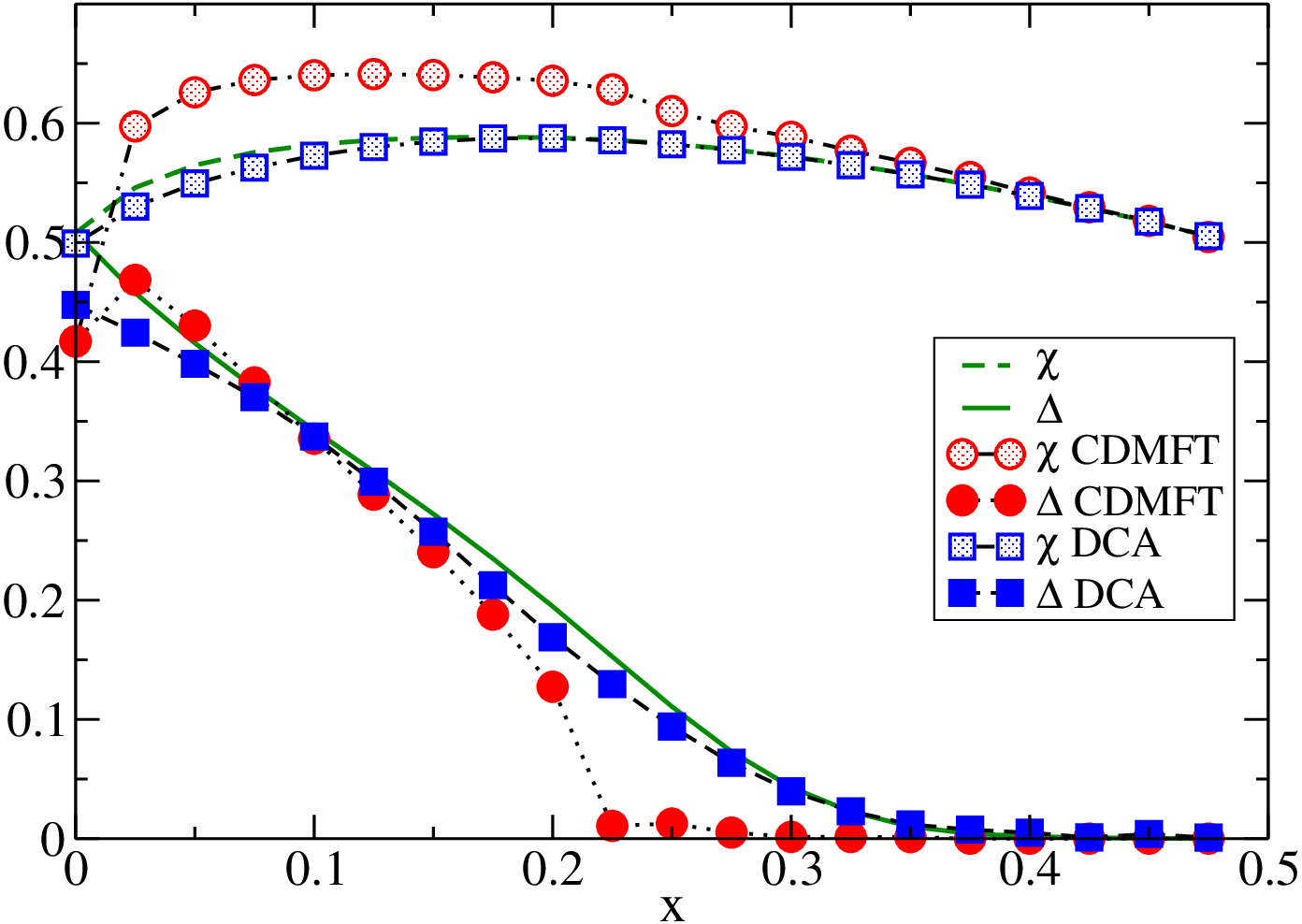}
\vskip 0.4cm
\includegraphics[width=4.5cm]{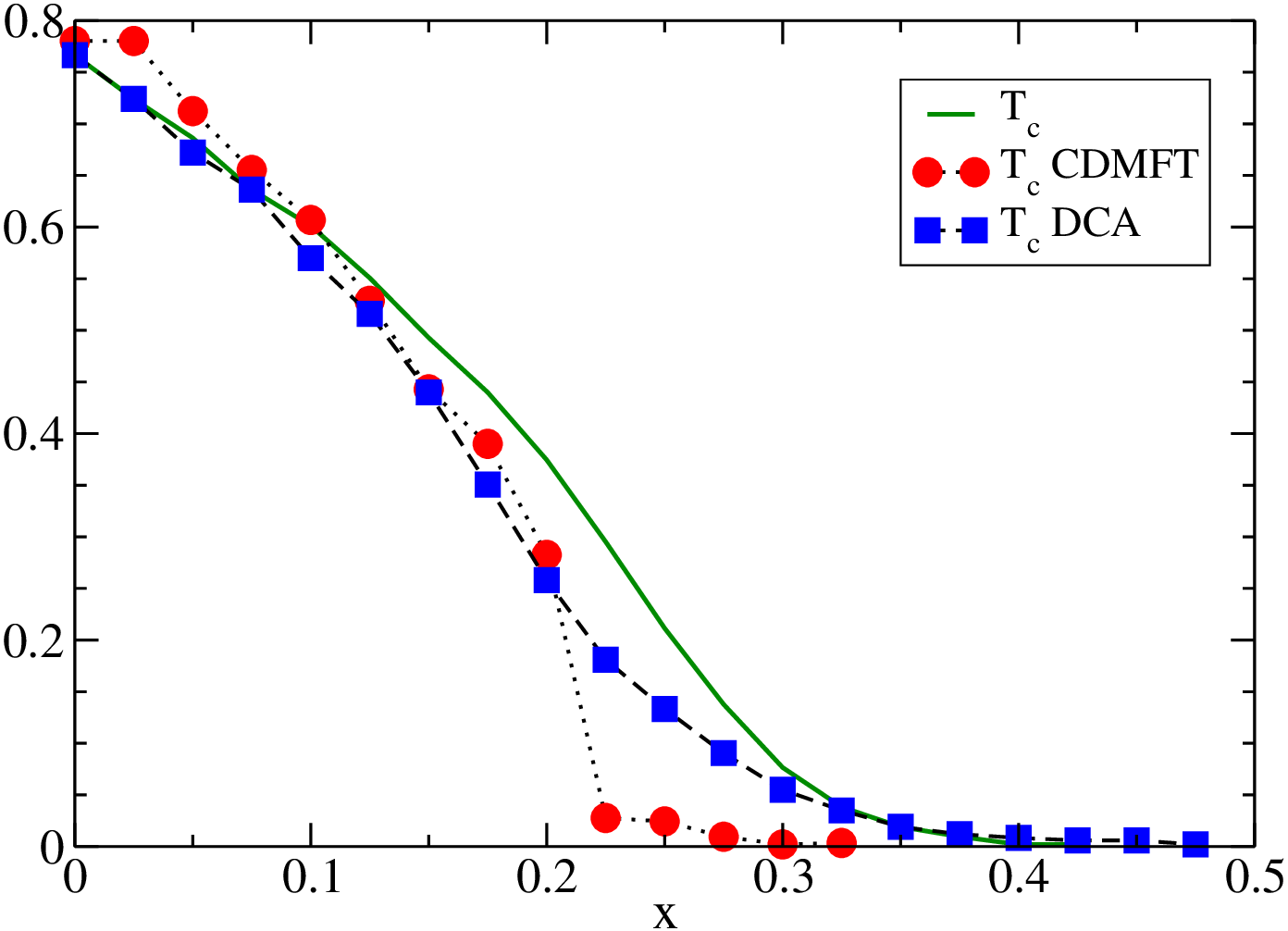}
\caption{(Color online) Same as Fig. \ref{Lc2} for $5\times 5$ cluster.
}\label{Lc5}
\end{figure}

We start with the $L_c=2$ system, namely the so-called $2\times 2$ plaquette. This is the smallest cluster that can host $d$-wave superconductivity, and its limited size makes it definitely the most popular system in CDMFT and DCA\cite{plaquette,dwaveclusters}.
From the results shown in Fig. \ref{Lc2} we find that both methods overestimate the $d$-wave order parameter, with stronger deviations at small $x$ for CDMFT and for larger $x$ in DCA. As far as the critical temperature is concerned, however, CDMFT turns out to reliably estimate the thermodynamical limit over almost the entire doping range, while DCA leads to a huge overestimate of $T_c$ (a factor of 2 at half-filling, which increases as the doping grows). 
These results suggest that the geometrical constraints imposed by the $2\times 2$ cluster have extremely strong effects on the $d$-wave phase, making this kind of cluster hardly useful for a quantitative estimate. Nonetheless, the simplicity of this cluster makes it a simple instrument to analyze the essential physics of two-dimensional correlated models.

As soon as we increase the size of the cluster to $L_c=3$, CDFMT experiences a substantial improvement. Both $\Delta$ and $T_c$ are indeed reasonably close to the exact solution, except for a moderate bifurcation of $T_c$ for large doping. 
Conversely, DCA strongly underestimates both quantities. It should be noted that odd values of $L_c$ explicitly break the particle-hole symmetry which holds at half-filling in the original model, since the $(\pi,\pi)$ point of the Brillouin zone is not included among the cluster momenta: this is the reason
why $\chi$ and $\Delta$ are not equal for $x=0$.
The success of CDMFT for this small cluster has a twofold interest: on one hand it is a possible promising direction for full numerical solutions of the Hubbard model, since the size of this cluster is reasonably small to allow for a reasonably accurate numerical accuracy; on the other hand it suggests us that, from a geometrical point of view, it is important to have at least two independent local amplitudes for the $\Delta$ field (the symmetry group of square clusters allows two independent bonds for $L_c=3$, in CDMFT, while in the corresponding DCA cluster there is only one independent $\Delta(\kk_c)$).

For the $L_c=4$ cluster the two approaches give essentially analogous results, and none of them is particularly interesting. This underlines the fact that, for such small values, the precise shape of the cluster matters more than the total number of sites, and that the inclusion of the $(\pi,\pi)$ point improves the DCA results, even if not by a large amount. 

Interestingly, at $L_c=5$ DCA becomes substantially more accurate than CDMFT. This confirms that DCA enters the asymptotic regime more rapidly than CDMFT, for which the finite-size effects survive to larger clusters, as we anticipated in the previous section.

\begin{figure}[h]
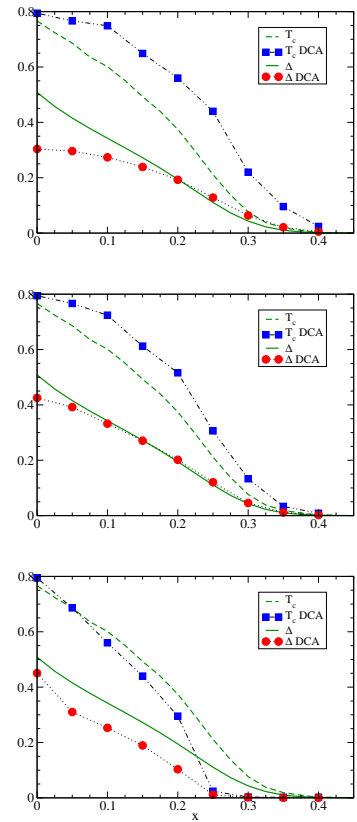

\includegraphics[width=4.5cm]{fig7a.eps}
\vskip 0.4cm
\includegraphics[width=4.5cm]{fig7b.eps}
\vskip 0.4cm
\includegraphics[width=4.5cm]{fig7c.eps}
\caption{(Color online) Doping dependence of $\Delta$ and $T_c$ in DCA for tilted clusters ($N_c=8, 12, 16$ from top to bottom).
}\label{tilted}
\end{figure}

\subsection{Tilted Clusters}
As we mentioned above, when dealing with very small clusters the DCA solution may suffer of severe size effects, associated to the presence or absence of characteristic cluster momenta, of special relevance, such as $(\pi,\pi)$, $(\pi,0)$, $(0,\pi)$.
A potential solution for this kind of sensitivity to the specific size
and shape of the cluster is the use of specific ``tilted'' lattices, as
shown in Ref.\onlinecite{tilted_DCA,tilted2}. These clusters are compatible with the space group of the lattice, and at the same time are expected to display less important size effects than the standard square systems.
Our analysis shows that, unfortunately, the improvement brought by the tilted lattices is not substantial. 
In Fig. \ref{tilted} we display DCA results for $N_c=8, 12, 16$ sites in tilted lattices, and we found that an accuracy comparable to that of the 
$5\times 5$ square cluster is obtained only for $N_c=20$, i.e., with almost the same number of sites.

\subsection{More specific lattices}
\begin{figure}[h]
\includegraphics[width=6cm]{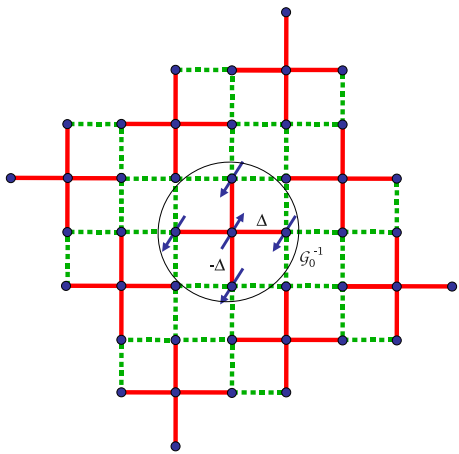}\caption{(Color online) Cross cluster and its embedding in the 2-d lattice.}\label{cross_cartoon}
\end{figure}

One of the outcomes of our analysis so far is that, as long as the size of the cluster is not sufficiently large, the accuracy of the results is dominated by geometrical factors. Therefore it is interesting to consider small specific clusters, whose size can be accessible to full numerical solutions, and that can minimize the geometrical frustration (or enhancement) of the $d$-wave superconducting state. 

To this aim we studied the ``cross'' cluster, shown in Fig. \ref{cross_cartoon} together with its embedding in the two-dimensional space, and small rectangular lattices. The star geometry can be considered as a good choice, since it can fit a $d$-wave ``cross'' of nearest-neighbor bonds. We find that the two approaches perform quite differently for this lattice. 
While CDMFT does not provide particularly accurate results, DCA reproduces remarkably well the exact solution for both the order parameter $\Delta$ and the critical temperature $T_c$. Interestingly, the accuracy in the superconducting parameters is not accompanied by an equally good description of the normal self-energy $\chi$ (results are not shown). The accuracy of DCA is associated to the particular values of the cluster momenta $K_c = (0,0),\, \pm ( 2\pi/5, 4\pi/5 )$ and  $\pm ( 4\pi/5, -2\pi/5 )$, which exclude the special symmetry points $(0,\pi)$ and ($\pi,0)$, but are at the same time close enough to the anti-nodal points to properly treat the superconducting order parameter. On the other hand, the momenta which are most important to describe $\chi$ are not included, leading to a worse estimate.

\vskip 0.4cm
\begin{figure}[h]
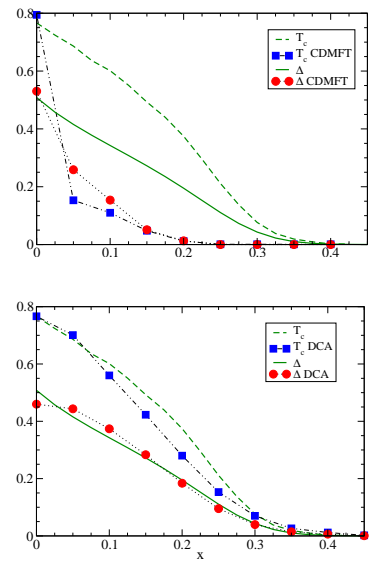

\includegraphics[width=4.7cm]{fig9a.eps}
\vskip 0.4cm
\includegraphics[width=4.7cm]{fig9b.eps}
\caption{(Color online) Doping dependence of $\Delta$ and $T_c$ for the cross cluster ($N_c=5$). Top panel is CDMFT, bottom panel DCA.
}\label{cross}
\end{figure}

\begin{figure}[h]
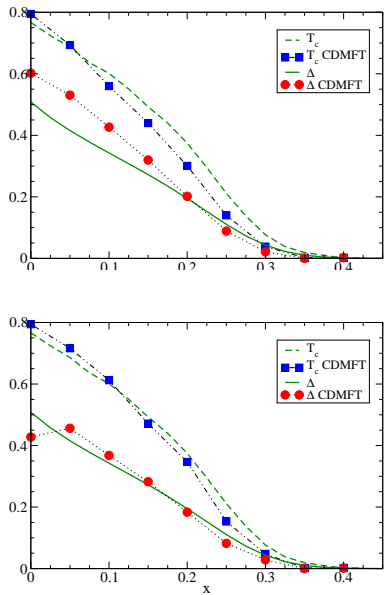

\includegraphics[width=5cm]{fig10a.eps}
\vskip 0.4cm
\includegraphics[width=5cm]{fig10b.eps}
\caption{(Color online) Doping dependence of $\Delta$ and $T_c$ in CDMFT for rectangular clusters ($N_c=6$ and $12$ from top to bottom).}\label{rectangles}
\end{figure}

As far as CDMFT is concerned, much closer agreement with the thermodynamic limit is reached, as shown in Fig. \ref{rectangles}, using (small) rectangular lattices.
Already the $2\times 3$ rectangle provides quite accurate results over the entire range of dopings, and the $3\times 4$ one shows a remarkable agreement with the thermodynamic limit. An explanation for the success of such clusters could be found in the relatively large number of independent bonds compared to their total number, which is a consequence of the lower symmetry of the cluster space group.
DCA on these rectangular lattices does not lead to particularly accurate results. Essentially the results (not shown) can be seen as a slight improvement on the corresponding square lattice (the largest square lattice contained in the rectangle), as far as $T_c$ is concerned.  

We have thus identified at least two relatively small lattices which provide accurate results and  that can be reasonably approached using a full numerical solution of CDMFT or DCA for the quantum Hubbard or t-J models, namely the 5-site cross for DCA and the 6-site rectangle for CDMFT. If we assume that dynamical effects will not spoil the geometrical effects that we have identified, these clusters could be an ideal compromise between accuracy and computational effort.

\section{conclusions}
In this paper we have investigated the accuracy of two cluster extensions of DMFT which are based on diametrically opposite perspectives: DCA, which enforces a momentum-space point of view, and CDMFT, which is formulated in real space. Even if the two methods are similar in spirit, they turn out to have different strengths and weaknesses. In this paper we analyzed the behavior of the two methods for a number of different clusters, using as a reference model an exactly solvable one which presents $d$-wave superconductivity as well as strong-correlation effects leading to the half-filling Mott physics.
The model under consideration is the two-dimensional t-J model treated within the mean-field slave-boson method. Even if this model is more naturally seen as an approximation of the real t-J model, here it is used to benchmark the cluster methods against an exact solution containing the essential physics of the cuprates.

The choice of the model does not allow us to discuss the frequency dependencies of the observables. 
Therefore our strategy is to focus on the geometrical effects introduced by the finite-size of the clusters 
(and eventually by their shape) within the different methods.

Analyzing the results as a function of the cluster size for square
lattices of linear size $L_c$, we found that the smallest cluster ($L_c=2$) 
provides rather inaccurate results, at least quantitatively.
This suggests that full CDMFT and DCA studies of two-dimensional models
could be poorly representative of the thermodynamic limit.
The evolution increasing the cluster size is quite irregular, but it
shows some important properties: CDMFT is found to adapt rather well
to some precise shapes, while DCA shows larger geometrical effects,
even though it converges faster to the thermodynamic limit (the latter is 
essentially reached for lattices of the order of
$5\times 5$ sites).
The limitations of DCA for small clusters are not dramatically reduced
by using ``tilted'' lattices, which include the most relevant momenta. 

On the other hand, it is found that specific small lattices can provide
very accurate results. In particular rectangular lattices provide 
rather accurate results within the CDMFT approach, even for the smallest
case of a $2\times 3$ rectangle. 
As far as DCA is concerned, we find that a 5-sites ``cross" cluster gives extremely 
accurate results for both the superconducting order parameter and the
critical temperature, even if the description of the normal self-energy is not
equally accurate. 
These two clusters (6-site rectangle in CDMFT and 5-site cross in DCA) 
provide probably the best compromise between size of the cluster
and computational cost, and they can be useful suggestions for future
full solutions of the actual two-dimensional systems such as the Hubbard and
the t-J models.

\begin{acknowledgements}
We acknowledge useful discussions with C. Castellani.
\end{acknowledgements}


\begin{thebibliography}{99}
\bibitem{notacenter} For even
values of $L_c$ the center of the cluster is not represented by a
single site, but instead by a 4--site plaquette; this sites are
however equivalent to each other, since they are connected by the
cluster symmetry group.
\end{thebibliography}
\end{document}